\documentclass[aps,prl,preprint,superscriptaddress]{revtex4-2}

\usepackage{amsmath}
\usepackage{amssymb}
\usepackage{graphicx}
\usepackage[font=small]{caption}
\usepackage{subcaption}
\usepackage{float}

\begin{document}

\title{Fermi Liquid Theory Sheds Light on “Hot” EHL in 1L-MoS\textsubscript{2}}

\author{R. L. Wilmington}
\affiliation{Department of Physics, North Carolina State University,
Raleigh, North Carolina 27695-8202, USA}
\affiliation{Naval Information Warfare Center Atlantic, 
Hanahan, SC 29419, USA}

\author{H. Ardekani}
\affiliation{Department of Physics, North Carolina State University,
Raleigh, North Carolina 27695-8202, USA}

\author{A. Rustagi}
\affiliation{Department of Physics, North Carolina State University,
Raleigh, North Carolina 27695-8202, USA}
\affiliation{School of Electrical and Computer Engineering, Purdue University,
West Lafayette, IN-47906}

\author{A. Bataller}
\affiliation{Department of Nuclear Engineering, North Carolina State University,
Raleigh, North Carolina 27695-8202, USA}

\author{A. F. Kemper}
\affiliation{Department of Physics, North Carolina State University,
Raleigh, North Carolina 27695-8202, USA}

\author{R. A. Younts}
\affiliation{Naval Information Warfare Center Atlantic, 
Hanahan, SC 29419, USA}

\author{K. Gundogdu}
\email[]{kgundog@ncsu.edu}
\affiliation{Department of Physics, North Carolina State University,
Raleigh, North Carolina 27695-8202, USA}

\date{\today}

\begin{abstract}

2D transition metal dichalcogenides (TMDCs) exhibit an electron-hole liquid phase transition at unusually high temperatures. Because these materials are atomically thin, optical excitation leads to material expansion. As a result, during the EHL phase transition the electronic band structure evolves due to both material thermal expansion and renormalization of the bands under high excitation densities. Specifically, these effects lead to indirect gap electronic band structure with a valence band maximum located at the $\Gamma$ valley. In this work we developed a methodology for analyzing the spectral evolution of the photoluminescence of suspended 1L-MoS\textsubscript{2} during the EHL phase transition by using Fermi liquid theory. The resulting analysis reveals valley-specific carrier densities, radiative recombination efficiencies, and intraband carrier relaxation kinetics in 1L-MoS\textsubscript{2}. More broadly, the results outline a methodology for predicting critical EHL parameters, shedding light onto the EHL phase transition in 2D TDMCs.
\end{abstract}

\maketitle

Tunability of optical and electronic response is the primary property of semiconductors that enables widespread utilization in various technological applications. Electrical conductivity, optical absorption, and emission are largely determined by the density of electrons and holes that can be created using chemical or optical doping. Often, the Coulomb and exchange interactions among these charge carriers can lead to bound exciton formation. At high concentrations, excitons form a plasma and then condense into an electron-hole liquid (EHL) state, which is a Coulomb-bound macroscopic phase of matter observable in many 2D and 3D semiconductors \cite{Keldysh1968,Sibeldin2017}. However, unique conditions such as cryogenic temperatures and specific electronic band structure are often required to form an EHL state. Our primary understanding of this phase relies on the early theoretical work of Keldysh, who first predicted an EHL phase transition in semiconductors in 1968 \cite{Keldysh1968}. Since then, experimental and theoretical studies on conventional semiconductors have been performed mainly at cryogenic temperatures. These studies create a framework for predicting EHL formation in new materials, due to the empirical relationship found between the exciton binding energy and the critical temperature below which EHL forms \cite{Keldysh1986}.

Recently emerging 2D materials have introduced a new paradigm for studying such high-density phases of matter \cite{Sibeldin2017}. Due to low dimensionality, and therefore inhibited screening, exciton binding in these materials is strong, making them good candidates to investigate the EHL phase at high temperatures \cite{Choi2015}. In the last few years, EHL has been experimentally confirmed in suspended monolayer (1L) MoS\textsubscript{2} above room temperature \cite{Younts2019}. These experiments were further supported by the completion of a theoretical phase diagram for 1L-MoS\textsubscript{2}, which shows a possible transition to an EHL phase for carrier densities exceeding $\sim 10^{12}$ cm$^{-2}$ \cite{Rustagi2018}. EHL formation in MoS\textsubscript{2} is evidenced by the changes in the photoluminescence (PL) spectra, which demonstrate abrupt intensity increase and spectral shift coincident with reaching a certain excitation threshold. Raman experiments combined with band structure calculations indicate that at these excitation densities, the material undergoes a direct to indirect bandgap transition \cite{Bataller2019}. As such, this phase transition involves not only carriers in the direct K valley as seen in unstrained 1L-MoS\textsubscript{2}, but also carriers in the indirect $\Gamma$ valley. Carrier densities in different valleys are therefore dependent not only on laser fluence, but also on the evolution of the electronic band structure due to thermal expansion. By using Fermi liquid theory, DFT band structure calculations, and Raman spectroscopy measurements, we have assembled a methodology for modeling the PL of the EHL phase of 1L-MoS\textsubscript{2}, and fit that model to experimental PL spectra taken at many optical excitation levels. From this analysis, we calculate the densities of carriers hosted in the K and $\Gamma$ valleys during the EHL phase transition. These results show that as the material reaches the highest possible excitation density, most of the holes reside in the $\Gamma$ valley. Surprisingly, the small population of holes in the direct K valley are responsible for the enhanced PL intensity. Also, intraband transitions resulting from the degeneracy of the system were shown to lead to a large spectral broadening of the electron/hole distribution of carriers, resulting in a broad PL peak with a low-energy tail.

\begin{figure}[ht]
	\includegraphics[width=0.5\linewidth]{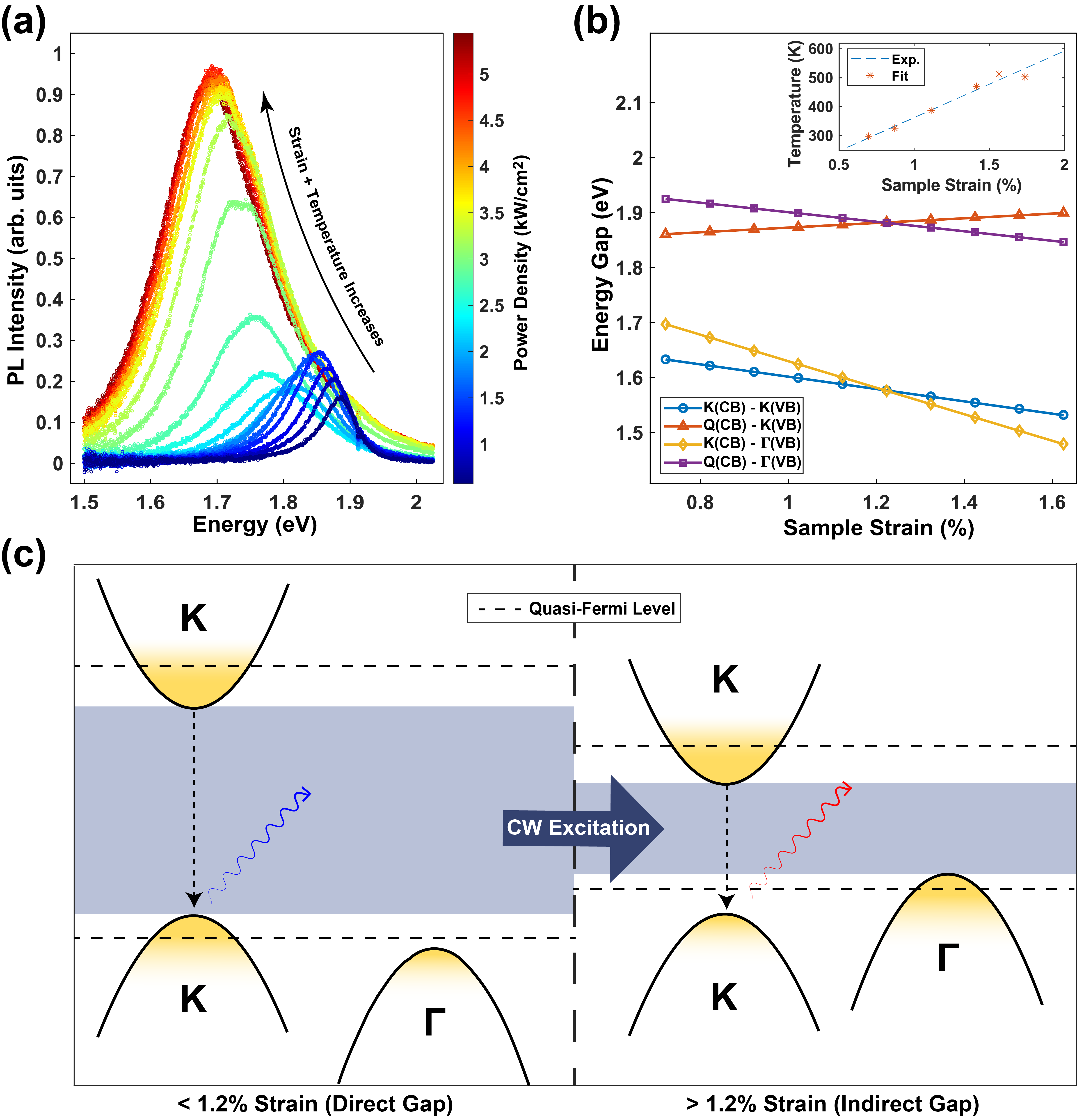}
	\caption{(a) PL measurements showing excitonic to EHL transition in suspended 1L-MoS\textsubscript{2} \cite{Bataller2019}. Increased fluence corresponds to increased strain and temperature, PL redshift and peak intensity increase. (b) Calculated bandgap shifts due to sample strain (referenced to K-VB). Inset shows fit of strain vs temperature based on Raman spectroscopy measurements \cite{Bataller2019}. (c) Schematic of band structure evolution during lattice expansion. Dashed lines indicate quasi-Fermi levels for electrons/holes. Shaded area shows the bandgap before and after phase transition.}
	\label{fig:1}
\end{figure}

As stated, the primary evidence for EHL phase transition in suspended 1L-MoS\textsubscript{2} is the evolution of the PL spectra with increasing fluence, shown in Fig. \ref{fig:1}(a) \cite{Bataller2019}. As incident continuous wave (CW) laser fluence increases from 1 to 5 kW cm$^{-2}$ the spectra first redshifts and decreases in amplitude. Then, at a threshold fluence with only a fraction of increase in the excitation density, a 5-fold increase in PL peak intensity is observed. These changes in the PL spectra are both due to photoexcitation-induced thermal expansion of the sample as well as the phase transition of free excitons (FE) into EHL. In order to determine the amount of strain and temperature increase, we performed Raman spectroscopy measurements at multiple fluences \cite{Bataller2019}. We then calculated the corresponding electronic band structure (see Supplemental Material \cite{Supplemental}). While unstrained, 1L-MoS\textsubscript{2} is a direct gap semiconductor at the K point \cite{Mak2010,Splendiani2010}. As laser heating brings the sample to 410\ K, the resulting 1.2\% strain causes the bandgap to become indirect [Fig. \ref{fig:1}(b, inset)]. Therefore, during the phase transition most of the excited holes reside in the $\Gamma$ valley. To quantitatively explore this phase transition, we developed a Fermi liquid theory model to predict the PL response of suspended MoS\textsubscript{2} monolayers under high excitation. 

The general form for the PL spectrum of EHL can be written as the convolution integral of all occupied conduction and valence band states as follows:

\begin{equation} \label{PL Intensity}
I(h\nu) = A |\mu|^2 \int_0^{h\nu'} D_e(E) D_h(h\nu' - E) dE,
\end{equation}

\begin{equation}
D_{e,h}(E) \sim g_{e,h} \cdot f_{e,h}(E),
\end{equation}
where $I(h\nu)$ is the spectral intensity and $D_{e,h}(E)$ is the distribution of carriers, given generally by the multiplication of the bare density of states ($g_{e,h}$) and a Fermi-Dirac distribution [$f_{e,h}(E)$]. Subscripts denote the charge species (electrons or holes) \cite{Pelant2012}. A is an overall scaling factor and $|\mu|^2$ is the matrix element of the quantum-mechanical transition for the direct gap photon emission event, treated as a fit parameter (see discussion below). For a 2D semiconductor, the bare density of states (DOS) is constant above/below the bandgap:

\begin{equation} \label{2D density of states}
g_{e,h} = \frac{\eta m_{e,h}^*}{\pi \hbar^2},
\end{equation}
where $\eta$ is the degeneracy of the band and $m^*$ is the corresponding effective mass. The upper integration limit ($h\nu'$) is related to the measured photon energy ($h\nu$) by the summation of all radiative flux contributions \cite{Pelant2012}. In this model, we consider the optical bandgap for unstrained 1L-MoS\textsubscript{2} ($E_{opt}$ = 1.89eV), the FE binding energy ($E_{bind}$ = 0.42eV) \cite{Ramasubramaniam2012,Yu2015}. Also included are the thermal bandgap reduction ($\Delta E_T$) and the bandgap renormalization energy ($E_{BGR}$):

\begin{equation} \label{Energy Offset}
h\nu' = h\nu - (E_{opt} + E_{bind} - \Delta E_T - E_{BGR})
\end{equation}
$\Delta E_T$ can be calculated by measuring relative shifts in the calculated band structure [Fig. \ref{fig:1}(c)] as a function of temperature (see Supplemental Material \cite{Supplemental}). $E_{BGR}$ is a correction to the energy bandgap resulting from high charge carrier density, supplied as a fitting parameter \cite{Schmitt-Rink1984}. Due to the large momentum gap between the K and $\Gamma$ valleys and the high concentration of direct K-K gap carriers, we neglect any contribution from phonon-mediated recombination. The electron/hole quasi-Fermi levels associated with a given PL intensity can be extracted from the sample temperature and charge carrier density (see Supplemental Material \cite{Supplemental}).

The general lineshape produced by Eq. \ref{PL Intensity} shows a high energy tail which closely resembles experimental PL measurements, but also a low energy tail which abruptly cuts off, failing to capture the nearly perfect exponential behavior experimentally observed \cite{Yu2019}. This broadening in the lower energy tail is due to finite lifetimes of the carriers within the degenerate bands \cite{Bagaev2010,Stoica2003}, i.e, vacancies in the electron/hole distribution can be filled by higher energy electrons/holes in the same band via inelastic collisions, transferring energy and momentum to other particles. The result is a finite lifetime of the final state of the radiative transition \cite{Martin1977,Snoke1990}. Accordingly, the distribution of carriers for electrons and holes are lifetime broadened, and can be modeled in the usual way with a Lorentzian convolution in energy:

\begin{equation} \label{Lorentzian Convolution}
D_{e,h}(E) = g_{e,h} \int_{0}^\infty f_{e,h}(E') \mathcal{L}_{e,h}(E - E') dE',
\end{equation}

\begin{equation} \label{Lorentzian Term}
\mathcal{L}_{e,h}(E - E') = \frac{1}{2\pi} \frac{\Gamma_{e,h}(E')}{\left(E - E'\right)^2 + \left(\frac{1}{2} \Gamma_{e,h}(E')\right)^2},
\end{equation}

\begin{equation} \label{Gamma factor_e}
\Gamma_{e,h}(E') = \alpha_{e,h}(E' - E_f^{e,h})^2 + C_{e,h},
\end{equation}
where $f_{e,h}(E)$ is the Fermi function for electrons/holes. Within Fermi liquid theory the inelastic collision rate, and therefore the broadening parameter $\Gamma_{e,h}(E)$, has a quadratic dependence on excitation energy \cite{Coleman2015}. Lifetime broadening was treated independently for electrons and holes, with $\alpha_{e,h}$ and $C_{e,h}$ serving as fitting parameters for each charge species (see Supplemental Material \cite{Supplemental}).

\begin{figure}[ht]
	\includegraphics[width=0.5\linewidth]{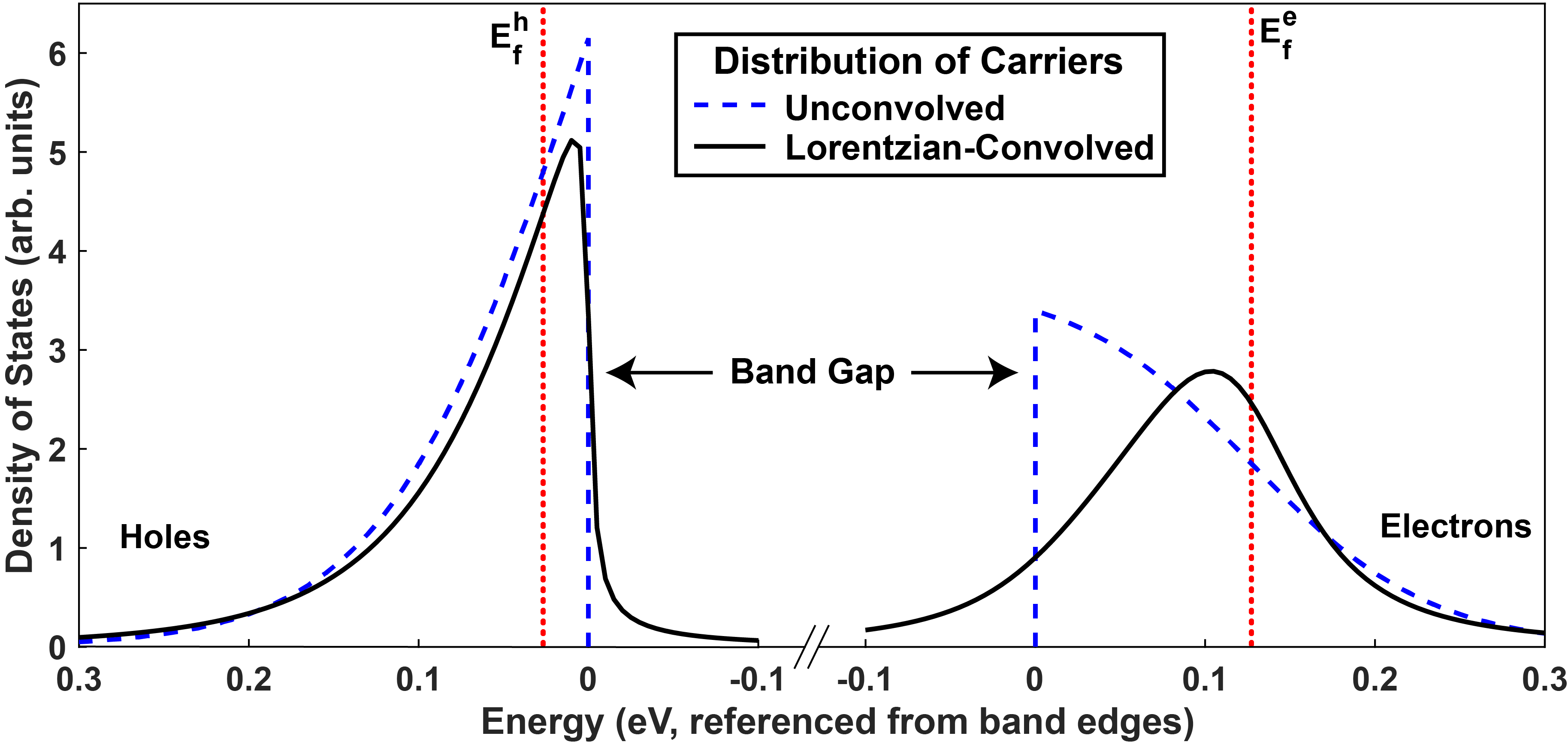}
	\caption{Distribution of carriers for holes (left) and electrons (right) referenced from $\Gamma-VB$ and $K-CB$, respectively. Shown with and without Lorentzian convolution. Red lines indicate quasi-Fermi levels after laser heating (T $\approx$ 600\ K). Note: bare DOS for 2D semiconductors are not energy-dependent, so the unconvolved distribution of carriers for electrons/holes (dashed lines) have purely Fermi-Dirac lineshapes.}
	\label{fig:2}
\end{figure}

Fig. \ref{fig:2} displays the broadened and unbroadened distribution of carriers for electrons and holes corresponding to the highest experimental excitation level of 1L-MoS\textsubscript{2}, based on the calculated quasi Fermi levels for each species. The energy scales for electrons/holes are referenced to conduction band K valley (K-CB) and the valence band $\Gamma$ valley ($\Gamma$-VB), respectively. The broadened distribution of carriers (solid black line) exhibits a significant deviation from the unconvolved distribution of carriers (dashed blue line). Fermi function weighting smooths the DOS away from the bandgap, but Lorentzian broadening smooths the DOS toward the bandgap as well. While the distribution of holes is only slightly modified, a larger broadening effect is observed for the distribution of electrons. When both populations are convolved, this results in both a high and low energy tail in the modeled PL spectra. The broadened distribution of carriers allows for some probability of optical transitions below the renormalized energy gap, better representing the observed PL lineshape. 

\begin{figure}[htb]
	\includegraphics[width=0.5\linewidth]{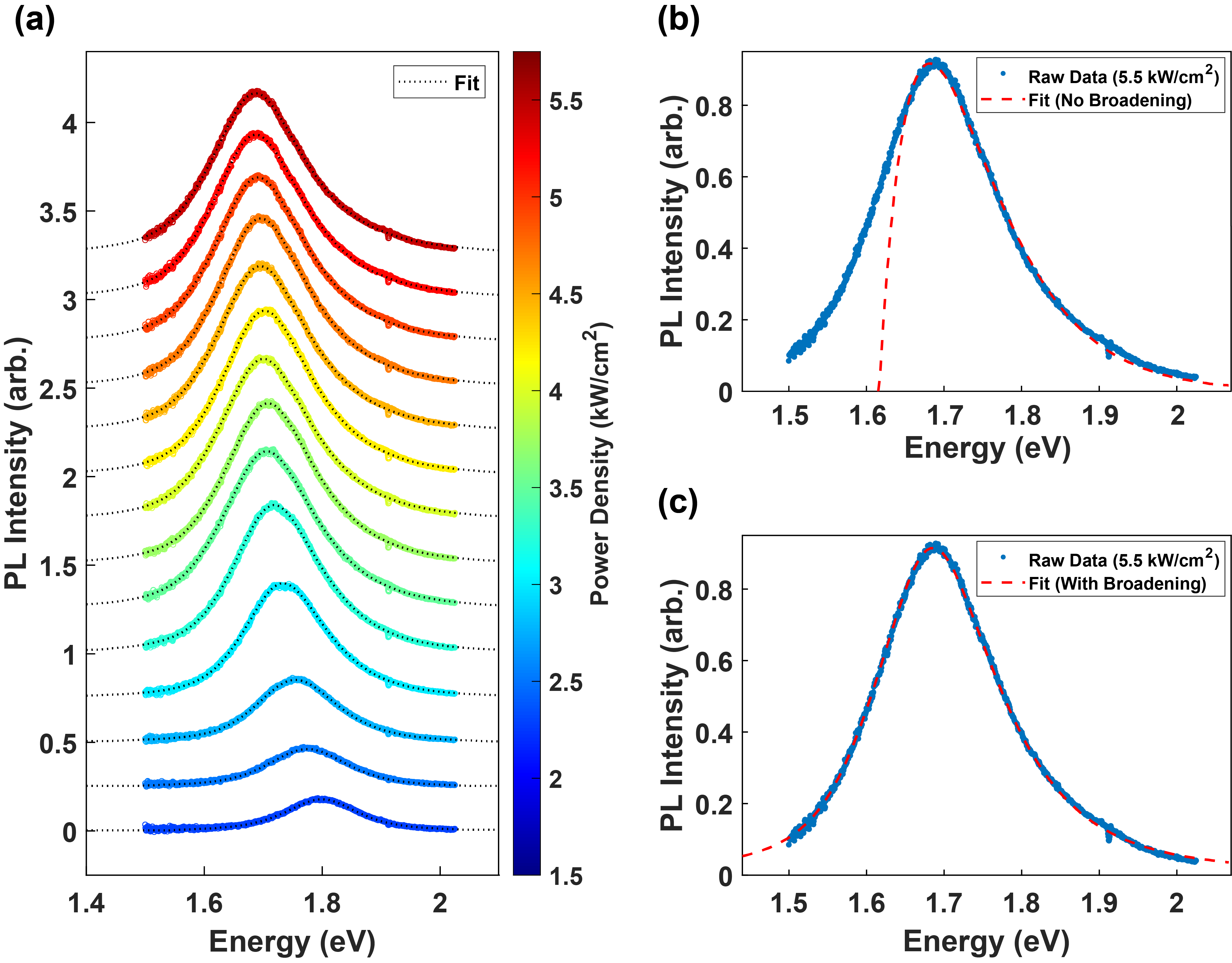}
	\caption{(a) Waterfall plot of model fits vs experimental spectra \cite{Bataller2019}. Fit shows good agreement with PL data beyond fluences of 2 kW cm$^{-2}$. Fit of highest fluence spectra calculated with (b) no intraband transition effects included, and (c) using distribution of carriers convolved with a Lorentzian to account for broadening due to intraband transitions.}
	\label{fig:3}
\end{figure}

\begin{figure}[htb]
	\includegraphics[width=0.25\linewidth]{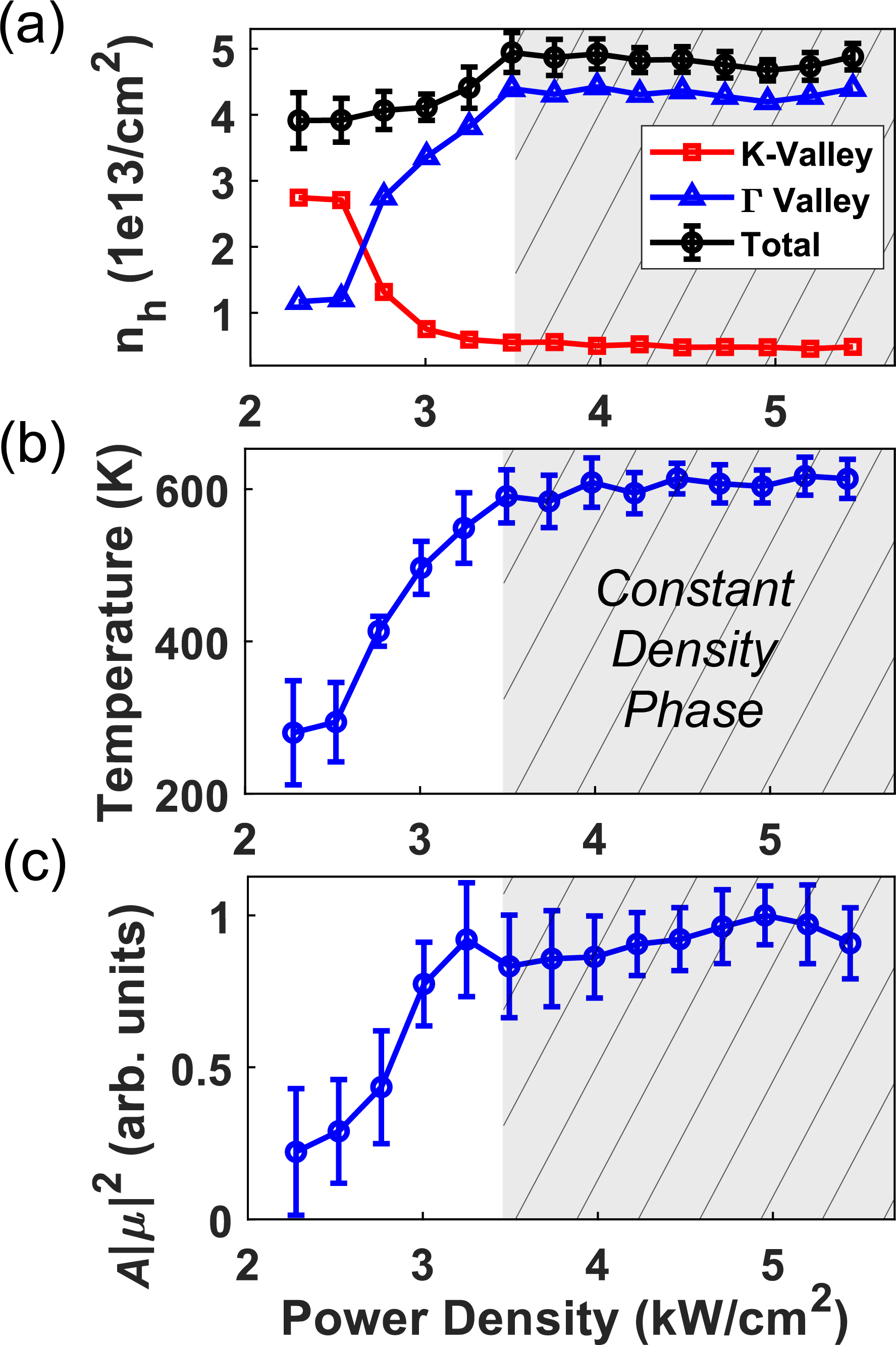}
	\caption{(a) Hole charge carrier density per valley ($n_h$), (b) calculated temperature and (c) PL intensity front factor plotted against CW laser power density. Saturation occurs at roughly 3.5 kW cm$^{-2}$, beyond which the system maintains a constant density EHL phase. Error bars indicate 2$\sigma$ calculated via Monte Carlo error estimation techniques (see Supplemental Material \cite{Supplemental}.)}
	\label{fig:4}
\end{figure}

By applying the appropriate energy convolution to the Lorentzian broadened distribution of carriers for electrons/holes (Eq. \ref{Lorentzian Convolution}), model PL outputs were produced and fit to experimental PL spectra. The fitted curves are plotted over the raw data in Fig. \ref{fig:3}(a). The EHL PL model produces a reasonable fit to the observed lineshape for the high fluence regime where the system is in the EHL phase, and well models the transition in the PL peak from approximately 2-3 kW cm$^{-2}$. The inclusion of the lifetime broadening effect discussed above greatly improved the model fit [Fig. \ref{fig:3}(b,c)]. However, for low fluences (below 2 kW cm$^{-2}$), the system is largely excitonic and our developed PL model begins to break down (see Supplemental Material \cite{Supplemental}). These fits provide critical information about the electronic properties of 1L-MoS\textsubscript{2} during the EHL phase transition. Fig. \ref{fig:4}(a) shows total charge carrier saturation near 3.5 kW cm$^{-2}$, reaching a peak charge carrier density between 4.5 and 5 $\times 10^{13}$ cm$^{-2}$. This agrees well with our previous estimates of the saturation carrier density at 4 $\times 10^{13}$ cm$^{-2}$ using the known relation between $E_{BGR}$ and carrier density (see Supplemental Material \cite{Supplemental}) \cite{Yu2019}. At the same fluence as the charge carrier density saturates, the temperature also appears to reach a maximum [Fig. \ref{fig:4}(b)], in agreement with previous Raman spectroscopy measurements \cite{Bataller2019}. The bandgap renormalization, dependent on the charge carrier density, also saturates around the same fluence. This corresponds to the qualitative observation that the PL peak has a constant redshift beyond fluences of 3.5 kW cm$^{-2}$, as shown in Fig. \ref{fig:1}(a). Above that fluence, the charge carrier density is saturated, as expected for the EHL phase. As a result, any further thermal bandgap reduction or bandgap renormalization is suppressed, and the PL emission peak stabilizes.

Using the fitted charge carrier density and temperature the strain, energy offset between the K and $\Gamma$ valleys, and quasi-Fermi levels for electrons and holes can be calculated. Combining these quantities allows the individual populations of charge carriers in each valley to be found, as shown in Fig. \ref{fig:4}(a) (for detailed calculation of quasi-Fermi levels and valley charge carrier densities, see Supplemental Material \cite{Supplemental}). For low fluences, when the bandgap is direct, holes exist mostly in the K valley. However, as the direct to indirect bandgap transition occurs, the hole population predominantly shifts to the $\Gamma$ valley. At high fluences, the saturation charge carrier density of the $\Gamma$ and K valleys are 0.48 and 4.40 $\times 10^{13}$ cm$^{-2}$, respectively, a ratio of 9:1. This finding suggests that despite the increase in PL intensity, the phase transition visible in the PL lineshape coincides with the migration of holes from the direct to the indirect gap, as the sample is moving from the excitonic to EHL phase.

Additionally, we are able to investigate the observed increase in PL intensity in the EHL phase. In the model for predicting PL response (Eq. \ref{PL Intensity}) the overall PL emission amplitude is determined by the prefactor $A|\mu|^2$, which contains both the magnitude of the transition dipole moment ($\mu$) for the direct bands and the relative density of electron hole pairs contributing to radiative and non-radiative recombination processes. The transition dipole moment can be approximated as independent of both energy and excitation level \cite{Coleman2015}. However, by allowing $A|\mu|^2$ to vary as a fit parameter over many fluences, we measure an approximately 4-fold increase [Fig. \ref{fig:4}(c)]. This is not simply a product of increasing charge carrier density; variation in the temperature and carrier concentration only account for $\approx 20\%$ of the PL emission increase seen in Fig. \ref{fig:1}(a). Noting that only direct gap carriers contribute to PL, we can normalize this value to the number of holes present in the K-valley. With that consideration, the PL intensity per unit of carrier density has a greater than 23-fold increase, indicating a massive change in the radiative decay dynamics of the EHL phase as compared to the excitonic regime (see further discussion below).

The model also provides insight into intraband scattering kinetics. The inclusion of Lorentzian broadening due to intraband transitions of charge carriers clearly improves the model fit, as shown in Fig. \ref{fig:3}(b,c), which compares the model fit with and without intraband scattering effects considered. With the Lorentzian convolution to the DOS, both the high and low energy tails match well to the observed PL. This indicates that a major contribution to the PL lineshape is from transitions of charges within the band filling vacancies left by other \textit{interband} transitions. This intraband scattering can be characterized by the inverse of the Lorentzian broadening parameter used to fit them, which is on the order of the intraband lifetime, $\Gamma \sim \hbar / \tau$. At saturation, the intraband lifetimes for electrons and holes are calculated to be roughly on the order of 10 fs (see Supplemental Material \cite{Supplemental}).

In summary, we have presented a theoretical analysis of the PL spectral lineshape of suspended 1L-MoS\textsubscript{2} as it undergoes EHL phase transition. Our results provide significant information regarding valley-specific carrier dynamics that help to quantitatively assess the changes in the optical characteristics of the material as it reaches the high density EHL phase. Strikingly, while most holes are in the $\Gamma$ valley, the PL efficiency of direct-gap (K-K) carriers increased 23-fold. There are two possible explanations: either radiative processes are increasing in efficiency, or nonradiative decay paths are being inhibited. It is unlikely that the 4-fold increase in PL amplitude can be attributed solely to changes in the transition matrix element, however suppression of nonradiative decay paths has been shown to significantly increase PL quantum yield \cite{Salehzadeh2014}. This is expected, as nonradiative processes (primarily defect assisted recombination, Auger recombination, and exciton-exciton annihilation) account for a majority of recombination events in 1L-MoS\textsubscript{2} \cite{Lien2019,Yu2016,sun2014observation}. Furthermore, it has been shown in previous studies that treatment of defects with molecular dopants can increase the PL efficiency of MoS\textsubscript{2} significantly \cite{ardekani2019reversible,amani2016recombination}. It is possible that the EHL phase transition provides sufficient carrier concentrations to fill defect sites, thereby increasing the recombination efficiency of direct gap carriers. Additionally, in many EHL systems, there is an enhancement of Auger processes that suppresses radiative recombination. However, for highly degenerate systems, conservation rules negate any first order Auger processes, and higher order processes like phonon-assisted Auger recombination dominate \cite{Haug1977}. We conclude that the observed 23-fold increase in the radiative recombination efficiency in the EHL phase is most likely due to suppression of defect-assisted recombination as well as first and higher order Auger processes.

Another open question is the precise nature of the evolution between the excitonic and EHL regime. To perform this PL experiment, sample heating and optical excitation both were supplied by a single CW laser. Therefore, the two variables cannot be decoupled. A traditional description of EHL condensation depicts free excitons slowly forming into larger multiexciton condensates until electron-hole droplet nucleation and growth begins \cite{Hanamura1977}. However, in this experiment exciton ionization and condensation processes are occurring simultaneously. It was predicted by Rustagi \textit{et al}. \cite{Rustagi2018} that the phase boundary between pure EHL, and EHL coexistent with FE gas, lies at carrier densities $\sim 10^{13}$ cm$^{-2}$, but that the carrier density threshold decreases by several orders of magnitude as the temperature increases above room temperature, up to the theoretical critical temperature of 515\ K \cite{Rustagi2018}. For these systems, increasing charge carrier density promotes the nucleation of EHL, and increasing temperature promotes the ionization of FE and the formation of EHP (electron-hole plasma). As laser power increases, it may be a combination of heat, carrier density, and the bandgap transition that collectively allow the EHL phase to occur.

These are just a few of the potential theoretical applications of our numerical results quantifying the carrier densities in each valley throughout the EHL transition. Modeling this transition with Fermi liquid theory and DFT band structure calculations produces results that agree with previous experimental measurements and align well with the phase transition regime predicted by theoretical phase diagrams for 2D MoS\textsubscript{2}. The incorporation of Lorentzian lifetime broadening due to intraband scattering for the degenerate system serves to well capture both the low and high energy side of the unique PL lineshape observed experimentally. However, the low-excitation non-degenerate excitonic phase of 1L-MoS\textsubscript{2} is not well described by these methods. The appearance of room temperature EHL states presents new and exciting opportunities for the use of 2D materials in high carrier density devices, or as a practical testbed for the dynamics of non-equilibrium steady states. This investigation deepens our understanding of the critical role of both interband and intraband charge dynamics in the formation of room temperature EHL phases, as well as their optical tunability.

\begin{acknowledgments}
The authors acknowledge funding from the National Science Foundation (NSF) grant DMR-1709934 and would like to thank the Naval Information Warfare Center Atlantic (NIWC-Atlantic) Naval Innovative Science and Engineering (NISE) program for funding the NIWC Atlantic portion of this research. A.F.K. was supported by NSF DMR-1752713. The views expressed in this article are those of the author(s) and do not necessarily represent the official position of the U.S. Navy. Distribution Statement A: Approved for Public Release. Distribution is unlimited (31Aug2020).
\end{acknowledgments}

\nocite{Giannozzi2009,Giannozzi2017,Perdew2008,QMEspresso,Monkhorst1976,Klingshirn1981}


%

\end{document}